\documentclass[a4paper,12pt]{article}
\usepackage[hmargin=.7in,vmargin=1.1in]{geometry}
\usepackage{indentfirst}
\usepackage{amsmath}
\usepackage{graphicx}
\usepackage{color}
\usepackage{authblk}
\usepackage{cite}
\usepackage{amsmath, amsfonts, amssymb}
\usepackage{graphicx}
\usepackage{float}
\usepackage{url}   
\usepackage{bm} 
\usepackage{upgreek}
\usepackage{mathtools}
\usepackage{multirow}
\usepackage{subfigure}
\usepackage{booktabs}
\usepackage{mathrsfs}
\usepackage{color,soul}

\newcommand\me{\mathrm{e}}
\newcommand\const{\text{const}}

\newcommand\pp{\uppi}
\newcommand{\dif}{\mathrm{d}}

\DeclareMathOperator{\diag}{diag}
\DeclareMathOperator{\Tr}{Tr}

\usepackage[bookmarksnumbered=true,bookmarksopen=true]{hyperref}
 \hypersetup{colorlinks,
             linkcolor=[rgb]{0,0.3,0.6}, 
             citecolor=[rgb]{0,0.3,0.6}, 
             urlcolor=[rgb]{0,0.3,0.6}}

\begin{document}

\title{\Large\textbf{Gliner Vacuum, Self-consistent Theory of Ruppeiner Geometry for Regular Black Holes}}

\author[1,2]{Chen Lan\thanks{lanchen@nankai.edu.cn} }
\author[1]{Yan-Gang Miao\thanks{Corresponding author, miaoyg@nankai.edu.cn}}

\affil[1]{\normalsize{\em School of Physics, Nankai University, 94 Weijin Road, Tianjin 300071, China}}
\affil[2]{\normalsize{\em Department of Physics, Yantai University, 30 Qingquan Road, Yantai 264005, China}}

\date{}

\maketitle

 \begin{abstract}
In the view of the Gliner vacuum, we remove the deformations in the first law of mechanics for regular black holes, 
where one part of deformations associated with black hole mass will be absorbed into enthalpy or internal energy, 
and the other part associated with parameters rather than mass will constitute a natural $V$-$P$ term.
The improved first law of mechanics redisplays its resemblance to the first law of thermodynamic systems,
which implies a restored correspondence of the mechanic variables to the thermodynamic ones.
In particular, the linear relation between the entropy and horizon area remains unchanged for regular black holes.
Based on the modified first law of thermodynamics, we establish a self-consistent theory of Ruppeiner geometry
and obtain a universal attractive property for the microstructure of regular black holes. 
In addition, the repulsive and attractive interactions inside and outside regular black holes are analyzed in detail.
 \end{abstract}

\tableofcontents

\section{Introduction}
\label{sec:intr}

It is closely related \cite{Ruppeiner:1995zz,Wei:2019uqg} to
the \emph{first law of thermodynamics} (1LT) and its corresponding entropy to construct the Ruppeiner geometry for black holes (BHs).
As is known, the 1LT is usually deduced from its resemblance to the \emph{first law of mechanics}  (1LM).
Based on $T\propto \kappa$
calculated from quantum theory \cite{Hawking:1974sw},
one can read from the resemblance the linear relation: $S\propto A$.
This is now known as the entropy/area law, where the corresponding entropy is dubbed as the Bekenstein-Hawking entropy (BHE) \cite{Bekenstein:1973ur,Hawking:1971tu}.
The 1LT cannot be correctly obtained when the resemblance mentioned above breaks.
Furthermore, if the entropy calculated from path-integral approach \cite{Gibbons:1976ue} or Wald's method \cite{Wald:1993nt,Brustein:2007jj}
does not coincide with that from the 1LT, the Ruppeiner geometry based on the 1LT will be unreliable.

A regular black hole (RBH) is such a system that its 1LM is deformed \cite{Zhang:2016ilt},
which brings about the breaking of the resemblance between its 1LM and 1LT.
From this point of view, a RBH does not have a well-defined 1LT,
and then its Ruppeiner geometry is suspect.
As the 1LT is considered to be the basis of Ruppeiner geometry and many other research topics,
such as the superradiance and area spectrum, etc.,
the lack of a well-defined 1LT leads to an obstacle for us to establish the Ruppeiner geometry for RBHs.

We take the noncommutative geometry inspired BH~\cite{Nicolini:2005vd} as an example of RBHs to show its present issue.
Its shape function, $f(r, M,\theta)$, contains two parameters, where
one is BH mass $M$ and the other $\theta$ is related to the minimal length of a noncommutative space.
If $\theta$ is set to be constant, the 1LT takes the form, $\dif M = T \dif  S$,
where $T$ is obtained from the Euclidean path integral near the horizon of this BH, see e.g. Ref.~\cite{Dabholkar:2012zz}.
The entropy calculated from such a 1LT breaks the entropy/area law, $S\neq A/4$.
On the other hand,  if the entropy of this BH without backreaction obeys \cite{Banerjee:2008gc} the entropy/area law, 
the 1LT does not holds, i.e., $\dif M \neq T \dif  S$, 
when the Hawking temperature is not modified and still calculated by the path-integral approach. 
In a word, this is the present situation for any RBHs\footnote{Their metrics satisfy $g_{tt}g_{rr}=-1$.} that either the entropy/area law or the 1LT breaks if the Hawking temperature is computed by the formula, $-g_{tt}^{\prime}(r_{\rm H})/{(4\pi)}$.

In this work we persist in adopting the Hawking temperature formula and maintain both the entropy/area law and the 1LT unbroken by applying the radial pressure of Gliner vacuum \cite{Gliner:1966,Ansoldi:2008jw} as RBH's pressure.
The so-called Gliner vacuum is an extended conception of vacuum.
It is introduced by Gliner and developed by Dynmnikova, Gurevich and Starobinsky, see the historical review \cite{Silbergleit2017} and references within.
The Gliner vacuum is different from the traditional vacuum with a vanishing energy-momentum tensor, $T_{\alpha\beta}=0$.
It is defined as a kind of matter that does not allow any preferred reference frame \cite{Dymnikova:1992ux}.
In the cases with spherical symmetry, this demands that $T^0_{\;\;0}=T^1_{\;\;1}$ \cite{Dymnikova:1992ux,Elizalde:2002yz} and provides an infinite set of comoving reference frames based on the Petrov's classification \cite{Petrov:2019}. Thus, the Gliner vacuum is anisotropic in spherically symmetric cases.

Due to the pressure of the Gliner vacuum, we can remove the deformations \cite{Ma:2014qma,Azreg-Ainou:2014twa} in the 1LM for RBHs.
In other words, the deformation related to $M$ will be absorbed into enthalpy or internal energy,
and the other deformations corresponding to the parameters rather than $M$  constitute a natural $V$-$P$ term.
For RBHs this procedure not only offers a solution to reconstruct the 1LT,
consequently a self-consistent theory of Ruppeiner geometry,
but also provides new insight into the 1LT related research topics, like the superradiance and area spectrum mentioned above.

This paper is organized as follows. In Sec.\ \ref{sec:regular-bhs}, we analyze the RBHs with spherical symmetry and a single shape function in terms of Ricci decomposition.
Next, we examine in Sec.\ \ref{sec:anomaly} the deformations of the first law of mechanics in the RBHs. We discuss in Sec.\ \ref{sec:pressure} how to remove the deformations in the first law of mechanics for some known RBHs in the view of the Gliner vacuum.
In Sec.\ \ref{sec:ruppeiner} we establish a self-consistent theory of Ruppeiner geometry for the RBHs with spherical symmetry and a single shape function by a well-defined 1LT.
The results show that all the RBHs are of attractive interaction.
It is known that the matters generating the RBHs violate the strong energy condition (SEC) around the center,
i.e., they have repulsive interactions.
Therefore, we try in Sec.\ \ref{sec:strong-energy} to explain in terms of the SEC on how the matters with repulsive interaction create the RBHs with attractive interaction.
To give a more intuitive illustration of the interaction structure of RBHs,  in Sec.\ \ref{sec:RN-RBH} we compare them with Reissner-Nordstr\"om black hole (RN BH) by considering the Raychaudhuri equation.
Finally, we give our summary in Sec.\ \ref{sec:summary}.

\section{Regular black holes}
\label{sec:regular-bhs}

We start with a spherically symmetric metric $g_{\mu\nu}=\diag\{-f(r),  f^{-1}(r),  r^2,  r^2\sin^2\theta\}$ with the shape function,
\begin{equation}
\label{eq:shape-function}
f(r)=1-\frac{2M}{r} \sigma\left(r, M, \alpha_i\right),
\end{equation}
where
$\alpha_i$'s are parameters rather than mass $M$. If $\sigma$ converges to one when $r$ goes to infinity,
the corresponding metric is asymptotic to the external spacetime of Schwarzschild BHs. In fact,
the asymptotic flatness of our metric only requires $\lim_{r\to\infty} \sigma/r = 0$,
i.e.,  $\sigma$ is divergent slower than $r$ as $r\to\infty$,
which is weaker than the condition of asymptotic to the Schwarzschild spacetime.
The requirement of asymptotic to the Schwarzschild BH simply guarantees the attractive nature (strong energy condition) and causality (dominant energy condition) outside the horizon.
In addition, for simplicity, we suppose that the BHs depicted by Eq.~(\ref{eq:shape-function}) are of two horizons at most.

It is convenient to apply the Ricci decomposition~\cite{Weinberg:1972kfs} for our investigations of RBHs,
\begin{equation}
\label{eq:ricci-decomp}
R_{\mu\nu\rho\sigma}=W_{\mu\nu\rho\sigma}+S_{\mu\nu\rho\sigma}+E_{\mu\nu\rho\sigma},
\end{equation}
where $W_{\mu\nu\rho\sigma}$ is the Weyl tensor, and
\begin{eqnarray}
S_{\alpha\beta\mu\nu}&=&\frac{R}{12}g_{\alpha[\mu} g_{\nu]\beta},\\
E_{\alpha\beta\mu\nu}&=&\frac{1}{2}\left(g_{\alpha[\mu} Z_{\nu]\beta}-g_{\beta[\mu} Z_{\nu]\alpha}\right),
\end{eqnarray}
with the traceless tensor,
\begin{equation}
Z_{\mu\nu}\equiv R_{\mu\nu}-\frac{1}{4}g_{\mu\nu} R.
\end{equation}
Thus, we rewrite $\sigma$ in terms of the curvature invariants,\footnote{In fact, $W$ and $E$ can be expressed by the three independent curvature invariants, $R$, $R_2$, and $K$, see App.\ \ref{app:shape-fun} for the details.} $R$, $W$, and $E$, when $r$ is around the center of RBHs,
\begin{equation}
\sigma= \frac{r^3 }{24 M} \left(R-2  \sqrt{3W}+3  \sqrt{2E}\right),\label{sigma1}
\end{equation}
where $W\equiv W_{\mu\nu\rho\sigma}W^{\mu\nu\rho\sigma}$ and $E\equiv E_{\mu\nu\rho\sigma}E^{\mu\nu\rho\sigma}$, and
both $W$ and $E$ are non-negative.
According to the relations,
 \begin{eqnarray}
R_2&\equiv& R_{\mu\nu}R^{\mu\nu}= \frac{1}{4}\left(R^2+2E\right),\\
K&\equiv& R_{\mu\nu\rho\sigma}R^{\mu\nu\rho\sigma}= \frac{1}{6}R^2+W+E,
\end{eqnarray}
the contraction of Ricci tensors $R_2$ and the Kretschmann scalar $K$ are also non-negative.
Based on the Ricci decomposition Eq.\ \eqref{eq:ricci-decomp}, the shape function then becomes
\begin{equation}
\label{eq:shape-new}
f(r)=1-\frac{\lambda(r)}{3}r^2,\qquad 4\lambda(r)\equiv R-2  \sqrt{3W}+3  \sqrt{2E}.
\end{equation}
That a BH is regular means that the
independent curvature invariants, $R$, $R_2$, and $K$, are convergent to finite constants everywhere in this BH spacetime.
In particular, as $r\to 0^+$ we have
\begin{equation}
R\sim R(0), \qquad
R_2\sim R_2(0), \qquad
K\sim K(0),
\end{equation}
with $R(0), R_2(0), K(0)=\const. <\infty$,
where the zero argument refers to the limit to the center of BHs.
We can see from Eqs.~(\ref{eq:shape-function}) and (\ref{sigma1}) that
the finiteness of  the shape function requests that $\sigma$ converges to zero faster than $r^3$, i.e.,
$\lim_{r\to0} \sigma/r^3=\const. <\infty$, see App.\ \ref{app:asym-sigma} for the detailed analyzes of $R$ and $K$.

It is worth mentioning that the treatment of RBHs is different from that of singular SBHs.
For the latter,  a complete Lagrangian is given, its equations of motion are solved, and thus the metric of SBHs is obtained.
For the former, however,
the first step is to propose such
a metric that the corresponding curvature invariants are finite
and geodesics are complete at the singularity of this metric;
and the second step is to construct the related Lagrangian that generates this RBH,
{\em i.e.}, to find out the reasonable matter source.

According to the behaviors of $\sigma$, the metrics of RBHs can be classified into two types.
The first type of RBHs is constrained locally by $0<\sigma\left(r, M, \alpha_i\right)\le 1$,
thus the outer horizon $r_+$ must be upper bounded by the Schwarzschild radius, $r_{\rm Sch}= 2M$, i.e.,
\begin{equation}
\label{eq:bound}
\frac{r_+}{r_{\rm Sch}}=\sigma\left(r_+, M, \alpha_i\right)\le 1.
\end{equation}
The second type of RBHs is picked out by the local condition, $\sigma\left(r, M, \alpha_i\right)> 1$. As the matter sources associated with the metrics in the second type are not clear,
we concentrate only on the first type and assume that $\sigma$ is a monotonically increasing function\footnote{The monotonicity comes from the positivity of energy density of matters, see Eq.~\eqref{eq:therm-variables}. Thus,  $\sigma(r)$ is a sigmoid function in the range of $r\in[0,\infty)$.} of $r$,  $\dif\sigma/\dif r>0$, in the remaining of the present work.

In addition, let us make a brief discussion about the factor $\lambda(r)$ in Eq.~(\ref{eq:shape-new}).
It is not difficult to check $\lambda(r)\sim R(0) /4$  in the limit of $r\to0^+$,
because $R\sim O(1)$, $E\sim O(r^2)$, and $W\sim O(r^2)$,
i.e. the orders of $E$ and $W$ are higher than that of $R$.
The same result can also be obtained by the method shown in Refs.~\cite{Melgarejo:2020mso,Bargueno:2020ais}, see App.\ \ref{app:asym-sigma}. Then changing the parameters (mass, charge, etc.) in $f(r_+)=0$,
such that $r_+$ approaches zero,
we obtain (see Eq.\ \eqref{eq:series-sigma})
\begin{equation}
1-\frac{R(0)}{12}r_+^2+O\left(r_+^3\right)=0,
\end{equation}
which implies that the horizon exists only for nonnegative $R(0)$.
Meanwhile, the negative $R$ is forbidden by the dominant energy condition \cite{Melgarejo:2020mso}.
In other words, a RHB is of a dS core rather than an AdS one around the center if $R(0)$ does not vanish.
Incidentally, if such a RHB is immersed in an AdS spacetime with the cosmological constant $\tilde\Lambda$, the corresponding shape function becomes
\begin{equation}
f(r)=1-\frac{\lambda(r)}{3} r^2-\frac{\tilde\Lambda}{3}r^2,\qquad
\tilde\Lambda<0,
\end{equation}
which gives rise to the fact that $\lambda(0)$ is larger than $-\tilde\Lambda$, otherwise, no horizons exist.

\section{Deformation of the first law of mechanics}
\label{sec:anomaly}

SBHs,  e.g., the Schwarzschild BH and Reissner-N\"ordstrom (RN) BH,
can be regarded as thermodynamic systems because their mechanic laws have
a resemblance to the thermodynamic ones.
Nevertheless, this resemblance is broken in RBHs.
If RBHs are of the shape function of Eq.~\eqref{eq:shape-function},
we make differentiation on the two sides of  $r_+=2M \sigma(r_+, M, \alpha_i)$ and obtain
\begin{equation}
\label{eq:dr}
\dif r_+=
\frac{2 \left(M   \sum_i\partial_{\alpha_i}\sigma \dif\alpha_i
+M  \partial_M\sigma  \dif M+ \sigma\dif M \right)}{1-2 M \partial_{r_+}\sigma }.
\end{equation}
To construct the 1LM, by substituting Eq.\ \eqref{eq:dr} into $\kappa \dif A/(8\pp)$,
where $A=4\pp r_+^2$ and $\kappa$ is surface gravity at $r_+$,
we give
\begin{equation}
\label{eq:def-1ML}
\frac{\kappa}{8\pp}\dif A=(1-\tau) \dif M
+\sum_i \beta_i \dif \alpha_i+\Delta(\alpha_i),
\end{equation}
where $\beta_i$ is the conjugate of $\alpha_i$, $\tau$ defined by
\begin{equation}
\tau \equiv 1-\frac{r_+}{2M}-M\frac{\partial \sigma(r_+, M, \alpha_i)}{\partial M}\label{taudef}
\end{equation}
means the deformation associated with mass $M$, and $\Delta(\alpha_i)$ stands for  the deformations associated with  $\alpha_i$'s.
To restore the resemblance between the mechanic and thermodynamic laws for RBHs, we have to remove reasonably those deformations just mentioned. In the following, we analyze the difficulties that we shall encounter and propose a possible way to overcome them.

\subsection{Deformations associated with $M$ and $\alpha_i$}

If one simply applies the traditional replacement for Eq.~\eqref{eq:def-1ML},
\begin{equation}
\label{eq:corresp}
\kappa \to 2\pp T,\qquad
A\to 4 S,\qquad
M\to E,
\end{equation}
one will obtain the formula,
\begin{equation}
\label{eq:def-1TL}
T\dif S=(1-\tau) \dif E
+\sum_i \beta_i \dif \alpha_i+\Delta(\alpha_i),
\end{equation}
which cannot be regarded as the 1TL of an isolated system because of the deformations.
In other words, either the RBH is not an isolated system or the correspondence between mechanic and thermodynamic variables Eq.\ \eqref{eq:corresp} is not appropriate.

The usual attempt is to let the term $\tau \dif M$ in Eq.~\eqref{eq:def-1ML} be absorbed into the entropy of RBHs to restore the resemblance between the mechanic and thermodynamic laws for RBHs. To this end, we define the  conditional entropy describing the entropy of RBHs,
\begin{eqnarray}
\label{eq:entropy-deviation}
S_{c}&\equiv&\int^{r_+}_{r_{\rm ext}} \frac{\dif M}{T} \nonumber \\
&=&\int_{r_{\rm ext}}^{r_+} \frac{\dif A}{4}+
\int_{r_{\rm ext}}^{r_+}
\frac{\tau}{T}\left(\frac{\dif M}{\dif \tilde r_+}\right)_{\alpha_i}\dif \tilde r_+,
\end{eqnarray}
where $T$ is Hawking temperature and $r_{\rm ext}$ denotes the horizon radius of extreme RBHs. In Eq.~\eqref{eq:entropy-deviation}, the first term is just the BHE, while the second one represents the deformation or deviation from the BHE.
We can prove that $\tau > 0$ when $0< \sigma\le 1$, which implies that the deformation in Eq.~\eqref{eq:entropy-deviation} is positive.
Let us analyze two aspects.
\begin{itemize}
\item If $\sigma$ does not contain $M$, our statement is obviously true due to Eq.~\eqref{eq:bound}, that is, $\tau > 0$ when $0< \sigma\le 1$.
\item If $\sigma$ depends on $M$ explicitly, we make a dimensionless rescaling  by $M$ for all variables, such as $r_+$ being rescaled to $x_+\equiv  r_+/(2M)$, thus the shape function becomes
\begin{equation}
f=1-\frac{\sigma(x, \widetilde \alpha_i)}{x},
\end{equation}
where $\widetilde \alpha_i$ is the dimensionless counterpart of $\alpha_i$ which is rescaled  by a power function of $M$.
Then repeating the procedure that proceeded at the beginning of this section, we obtain
\begin{equation}
\tau=1- x_+ \left[1 -\frac{\partial \sigma(x_+,\widetilde\alpha_i)}{\partial x_+}\right].
\end{equation}
Since the slope of tangent line of function $\sigma(x,\widetilde\alpha_i)$
at the outer horizon $x_+$ is not greater than unit when $0< \sigma\le 1$ and $x_+<1$, we verify the above statement.
\end{itemize}

If the BHE were replaced by the conditional entropy, we would remove the deformation associated with $M$
 but we would no longer maintain the linear relation between $S_c$ and the horizon area as a price. Furthermore,
the deformation with respect to $\alpha_i$ remains.
As a matter of fact, due to the lack of resemblance
between RBHs and the traditional thermodynamic systems, the introduction of $S_c$ does not work well in removing the deformations associated with $M$ and $\alpha_i$.
More importantly, $S_c$ is not an independent variable because its definition is subject to the 1TL.

\subsection{Proposal for removing deformations}

Starting with the classical action and observing the partition function at the zero-loop approximation,
we find that the Einstein-Hilbert action of RBHs with the shape function Eq.~(\ref{eq:shape-function})  contributes one part of entropy and the Gibbons-Hawking-York surface term provides the other part of contributions to the entropy of RBHs, and that the combination of the two contributions recovers the linear relation $S\propto A$. Let us give the derivation.  The Einstein-Hilbert action takes the form,
\begin{equation}
I_{\rm EH}=\frac{2 M-r_+ \left[1+2 M \sigma'\left(r_+\right)\right]}{4T},
\end{equation}
where $\sigma$ converges to unit as $r\to \infty$ and the prime denotes the derivative with respect to the radial coordinate, and the Gibbons-Hawking-York surface term is
\begin{equation}
I_{\rm GHY}=-\frac{M}{2T},
\end{equation}
where the flat spacetime is selected as background reference~\cite{Lan:2021klp}.
As a result, the total action reads
\begin{equation}
I_{\rm tot} =I_{\rm EH}+ I_{\rm GHY}=-\pp  r_+^2+\beta  \frac{r_+}{2}, \qquad \beta\equiv \frac{1}{T},
\end{equation}
with which the entropy can be computed by
\begin{equation}
S=\beta \frac{\partial I_{\rm tot}}{\partial \beta}-I_{\rm tot}=\pp r_+^2.
\end{equation}
That is to say, the semiclassical approach prefers that the entropy of RBHs is $A/4$ rather than  $S_{c}$.
The same result can also be obtained by the Wald method \cite{Lan:2021klp}.
The situation of RBHs is similar to the case of dielectric in an external electric field \cite{Landau:vol8} where the variation of internal energy does not alter the entropy of a dielectric system.
That is to say, the deformation associated with mass $M$ does not depend on the thermodynamic state of RBHs,
thus it should not affect the entropy of RBHs.

In conclusion, our proposal is to maintain the linear relation, $S\propto A$, and simultaneously to remove the deformations mentioned in the above subsection, that is, to establish the resemblance between the mechanic and thermodynamic laws for RBHs in terms of the pressure of  Gliner vacuum \cite{Gliner:1966,Ansoldi:2008jw} which is dealt with as the pressure of RBHs. Thanks to the pressure of the Gliner vacuum, the deformation associated with  $M$ will be absorbed into the enthalpy or internal energy and the deformation associated with $\alpha_i$ will constitute a naturally $V$-$P$ term. In this way, all deformations disappear and the other terms are well defined in Eq.~(\ref{eq:def-1ML}), which gives rise to a normal 1LT in the formulation.

We shall see that our proposal works well for some known RBH models as examples. It is necessary for us to choose models because we need the formulations of shape functions, however, we shall see that our proposal is valid for all RBHs with a single shape function depicted by Eq.~(\ref{eq:shape-function}).

\section{Pressure and reconstruction of  the first law of mechanics}
\label{sec:pressure}

It is widely known that the negative cosmological constant can be
interpreted~\cite{Kastor:2009wy} as thermal pressure of BHs,
while the positive cosmological constant has a problem of thermal equilibrium~\cite{Dolan:2010ha},
which is caused by the existence of a cosmological horizon.
No matter whether the cosmological constant is negative or positive,
the essence of this idea is to regard the pressure
of \emph{vacuum} as the pressure of BHs.
Nevertheless, a RBH has its own vacuum even though it is not involved in an additional AdS (or dS) term in metric.
This seed goes back to Sakharov and Gliner \cite{Sakharov:1966aja,Gliner:1966},
who re-explained the vacuum as spacetime filled with vacuum.
Such a vacuum is currently dubbed a Gliner vacuum.
It will be our key to reconstruct the 1LM for RBHs. 
Next, we introduce the pressure of the Gliner vacuum in the following two models.

\subsection{Model I}
\label{subsec:dymonikova}

The first model is generated~\cite{Dymnikova:1992ux} by an anisotropic vacuum and its shape function reads
\begin{equation}
f(r)=1-\frac{2M}{r} \sigma(r, M, \Lambda),\qquad
\sigma(r, M, \Lambda)=1-\exp\left(-\frac{\Lambda}{6M}r^3\right).
\end{equation}
Here $\Lambda$ is a cosmological constant.
The asymptotic flatness at infinity requires that $\Lambda$ be positive.
Moreover, $f(r)$ approaches to de Sitter space, i.e.,
$f(r)\sim 1-\Lambda r^2/3+O(r^5)$, as $r\to 0$.
When $r$ becomes large,
the term $\exp[-\Lambda r^3/(6M)]$ goes to zero,  
thus $f(r)$ is asymptotic to the Schwarzschild black hole.
Since $\sigma\le 1$ for $r\in \mathbb{R}^+$,
the outer horizon
$r_{\rm +}$ is restricted by $r_{\rm ext}\le r_{\rm +}\le 2M$
when $\Lambda  >9.28/(4M^2)$.
The extreme radius is $r_{\rm ext}\approx 0.85 \times 2M$.
For $\Lambda<0$, there is no real solution for $f(r_+)=0$.
Therefore, this model has no cosmological horizons, and thus does not suffer from the problem of thermal equilibrium.

The Smarr formula was obtained from
the total mass represented by the Komar integral
which can be separated \cite{Bardeen:1973gs} into two parts.
The first part is a surface integral over the horizon, and gives $\kappa A/(4\pp)$;
while the second one, including a deviation from the first law, is a volume integral with one boundary at spatial infinity and the other at event horizon \cite{Zhang:2016ilt,Gulin:2017ycu}. Combining the two parts, one obtains the Smarr formula,
\begin{equation}
\label{eq:smarr-dymnikova}
M=\frac{ \kappa A}{4 \pp }
+\epsilon M
+\frac{1}{2}\epsilon \Lambda  r_+^3,
\end{equation}
where $\epsilon\equiv\exp\left[-\Lambda  r_+^3/(6 M)\right]> 0$.
This Smarr formula suggests an extended phase space.
However, if one applied $P=-\Lambda/(8\pp)$ as thermal pressure \cite{Kastor:2009wy},
the first law of mechanics would be deformed, i.e.,
\begin{equation}
\label{eq:dymnikova-firstlaw1}
\frac{\kappa}{8\pp}\dif A= (1-\tau)\dif M
- \epsilon V\dif  P,
\end{equation}
where
$\tau \equiv\epsilon
\left[1-\ln \left(\epsilon\right)\right]$, and $V\equiv 4\pp r_+^3/3$ is the thermal volume inside the horizon.
Further, we find that
the $1$-form $\kappa \dif A /(8\pp) + V \dif P$ does not satisfy the integrable condition,
i.e., it cannot be written as a total derivative of any functions.

By introducing the radial pressure of the Gliner vacuum at the outer horizon,
\begin{equation}
P_{+}=\left.\frac{G^r_{\; r}}{8\pp}\right|_{r=r_{+}}=-\epsilon\frac{\Lambda}{8\pp}, 
\end{equation}
which is negative, as the pressure for Model I, where $G^r_{\; r}$ is $r$-$r$ component of Einstein tensor $G_{\mu\nu}$, we note that
the last term of  Eq.~\eqref{eq:smarr-dymnikova} is nothing else but $-3V P_{+}$. As a result, following the way in Ref.~\cite{Azreg-Ainou:2014twa} and introducing \emph{enthalpy},
\begin{equation}
H= (1-\tau)M,\label{entha1}
\end{equation}
we can reconstruct the 1LM from Eq.~(\ref{eq:dymnikova-firstlaw1}),
\begin{equation}
\label{eq:dymnikova-firstlaw2}
\frac{\kappa}{8\pp}\dif A= \dif H- V\dif  P_{+},
\end{equation}
and write the corresponding 1LT,
\begin{equation}
\dif H=T\dif S+ V\dif  P_{+},\label{lawthermo}
\end{equation}
with
\begin{equation}
T=\frac{\kappa}{2\pp}, \qquad S=\frac{A}{4}.
\end{equation}
The deformations of Eq.~\eqref{eq:dymnikova-firstlaw1} are completely removed in Eq.~\eqref{eq:dymnikova-firstlaw2}.

Alternatively, Eq.~\eqref{eq:dymnikova-firstlaw2} can be rewritten as
\begin{equation}
\frac{\kappa}{8\pp}\dif A= \dif U+ P_{+} \dif V,\label{flm}
\end{equation}
 with \emph{total internal energy}
 \begin{equation}
 U=(1-\epsilon) M=H-V P_+,
\end{equation}
which suggests that the deviation from $M$ in Eq.~\eqref{eq:smarr-dymnikova} should be absorbed into internal energy.
The other compelling reason to think of $(1-\epsilon) M$ as the energy is that it can be calculated by the integration of energy density $\rho$ over the whole space inside the RBH, i.e.,
\begin{equation}
\label{eq:internal-energy}
(1-\epsilon) M =\int^{r_{+}}_0 \int^{2\pp}_0\int^\pi_0\sqrt{-g} \rho\dif r \dif\phi\dif\theta=\frac{r_+}{2},
\end{equation}
where $\rho$ is defined by $\rho =-G^0_{\;0}/(8\pp)=\Lambda  \exp\left[-\Lambda  r^3/(6 M)\right]/(8\pp)$ according to Einstein's equation, $G^\mu_{\;\nu}=8\pp T^\mu_{\;\nu}$.
In other words,
$M$ is the total energy filled in the whole space of the RBH,
while $\epsilon M $ is the energy outside the RBH.

Thus, the internal energy of the RBH should be the energy enclosed in the event horizon, and the corresponding 1LT can be cast as follows:
\begin{equation}
\dif U = T\dif S -P_{+} \dif V,\label{fll}
\end{equation}
where every term is well-defined.
As we expect,  the resemblance between the mechanic and thermodynamic laws for this RBH is restored, see Eqs.~(\ref{eq:dymnikova-firstlaw2}) and (\ref{lawthermo}) or Eqs.~(\ref{flm}) and (\ref{fll}), and simultaneously the entropy is just the Bekenstein-Hawking entropy obtained by path-integral and Wald's method \cite{Lan:2021klp}.

\subsection{\label{subsec:ads-dymonikova}Model II}

Now let us turn to the second model, Model I immersed in AdS spacetime,
which can be established if the other cosmological constant $\widetilde\Lambda$ is introduced into Model I,
\begin{equation}
\label{eq:shape-ads-dymnokova}
f(r)=1-\frac{2M}{r} \sigma(r, M, \Lambda)-\frac{\widetilde\Lambda}{3}r^2,\qquad
\sigma(r, M, \Lambda)=1-\exp\left(-\frac{\Lambda}{6M}r^3\right).
\end{equation}
There is no doubt that such a metric is a solution of
Einstein's equations with a cosmological constant term \cite{Dymnikova:1992ux}.
The homogeneity of the universe remains unchanged
if $\Lambda$ is treated as a local character,
i.e., $\Lambda$  dominates only the inside (and around a certain range) of this RBH.
Meanwhile, to bypass the problem of thermal equilibrium \cite{Dymnikova:1998pm},
we demand $\widetilde\Lambda<0$.
The Smarr formula can be calculated in the same way as that mentioned above,
\begin{equation}
M=\frac{ \kappa A}{4 \pp }
+\epsilon M
+\frac{1}{2}\epsilon\Lambda  r_+^3
+\frac{1}{3}\widetilde\Lambda r_+^3,
\end{equation}
where
$\epsilon$ has the same form as that of Model I,
but the energy density now is calculated by $\rho=-(G^0_{\;0}+\widetilde\Lambda)/(8\pp)$
because of $G^\mu_{\;\nu} +\widetilde\Lambda g^\mu_{\;\nu}=8\pp T^\mu_{\;\nu}$.
We note that the deformations still exit
even if the AdS constant $\widetilde\Lambda$ is
regarded as pressure.
If we regard the radial pressure of the Gliner vacuum as the pressure,
we can obtain the same 1LM as Eq.~\eqref{eq:dymnikova-firstlaw2}, where the \emph{enthalpy} is defined in Eq.~(\ref{entha1}) and the Gliner pressure takes the form,
\begin{equation}
P_{+}=
-\frac{1}{8\pp}\left(
\epsilon \Lambda+\widetilde\Lambda\right).
\end{equation}
The competition of two cosmological constants appears in the pressure.
As we have noted in Sec.~\ref{sec:regular-bhs}, $\epsilon \Lambda$ should be larger than $-\widetilde\Lambda$, otherwise there will be no horizons.
As a result, the Gliner pressure in Model II is negative,
and this pressure is different from
the one simply introduced from the AdS cosmological constants.

\subsection{Other models}
\label{subsec:others}

The models in the above two subsections give a heuristic evidence
that a well-defined 1LM can be reconstructed without any deformations and the traditional area law, $S=A/4$, still holds when we introduce the radial pressure of Gliner vacuum.
It will be seen in this subsection that some other models,
including those with Gliner vacuum or vacuum-like mediums, also support this statement.

(i) Model in Ref.~\cite{Nicolini:2005vd}. This model is also called noncommutative geometry inspired black hole.
The shape function takes the form,
\begin{equation}
f(r)= 1-\frac{4M}{\sqrt{\pp}\,r}\gamma\left(\frac{3}{2},\frac{r^2}{4\theta}\right).
\end{equation}
The Smarr formula is
\begin{equation}
   M=
   \frac{A \kappa }{4 \pp }
   +\epsilon M
   +\frac{M r_+^3 \me^{-\frac{r_+^2}{4 \theta }}}{2 \sqrt{\pp } \theta ^{3/2}},
 \end{equation}
where in the second term on the right-hand side $\epsilon$ equals
\begin{equation}
\epsilon = \frac{2 }{\sqrt{\pp }}\Gamma \left(\frac{3}{2},\frac{r_+^2}{4 \theta }\right),
\end{equation}
which coincides with the result given by Eq.~\eqref{eq:internal-energy};
the third term corresponds to $-3V P_+$, and the radial pressure of the Gliner vacuum at the outer horizon reads
\begin{equation}
P_+=-\frac{M e^{-\frac{r_+^2}{4 \theta }}}{8 \pp ^{3/2} \theta ^{3/2}}.
\end{equation}
We can also give the 1LM Eq.~\eqref{eq:dymnikova-firstlaw2} with the enthalpy and internal energy,
\begin{subequations}
\begin{equation}
H
=\frac{r_+}{2}+
\frac{r_+^4}{12 \theta  r_+-12 \sqrt{\pp }\, \theta^{3/2} \exp\left({\frac{r_+^2}{4 \theta }}\right) \text{erf}\left(\frac{r_+}{2 \sqrt{\theta }}\right)},
\end{equation}
\begin{equation}
U=M \left[\text{erf}\left(\frac{r_+}{2 \sqrt{\theta }}\right)-\frac{r_+ \me^{-\frac{r_+^2}{4 \theta }}}{\sqrt{\pp } \sqrt{\theta }}\right]=\frac{r_+}{2}.
\end{equation}
\end{subequations}

(ii) Model in Ref. \cite{Hayward:2005gi}.
The shape function is
\begin{equation}
f(r)=1-\frac{2M r}{r^2+6M/(r\Lambda)},
\end{equation}
where $\Lambda$ is supposed to be positive.
The Smarr formula has the form,
\begin{equation}
M=\frac{\kappa A}{4 \pp }
+\frac{6 M^2}{\Lambda  r_{+}^3+6 M}
+\frac{18 \Lambda  M^2 r_{+}^3}{\left(\Lambda  r_{+}^3+6 M\right)^2},
\end{equation}
where the second term on the right-hand side stands for $\epsilon M$ with $\epsilon=6M/(6M+\Lambda r_{+}^3)$, which coincides with the result given by Eq.~\eqref{eq:internal-energy};
the third term corresponds to $-3V P_+$, and the radial pressure of the Gliner vacuum at the outer horizon equals
\begin{equation}
P_+=-\frac{9 \Lambda  M^2}{2 \pp  \left(6 M+\Lambda  r_{+}^3\right)^2}.
\end{equation}
The same procedure leads to the 1LM Eq.~\eqref{eq:dymnikova-firstlaw2} with the enthalpy and internal energy,
\begin{equation}
H=\frac{r_+^2}{4 M}, \qquad 
U=(1-\epsilon)M=\frac{r_+}{2}.
\end{equation}

(iii) Model in Ref.~\cite{Bardeen:1968non}. This model is usually called Bardeen's black hole.
The shape function is
\begin{equation}
f(r)=1-\frac{2M r^2}{\left(b ^2+r^2\right)^{3/2}},
\end{equation}
where $b$ is a magnetic charge.
The Smarr formula then reads
\begin{equation}
    M=\frac{A \kappa }{4 \pp }
    +M \left[1-\frac{r_{+}^3}{\left(b^2+r_{+}^2\right)^{3/2}}\right]
    +\frac{3 b^2 M r_{+}^3}{\left(b^2+r_{+}^2\right)^{5/2}},
\end{equation}
where we have used $\epsilon=1-r_{+}^3/\left(b^2+r_{+}^2\right)^{3/2}$. In this model, the pressure of the Gliner vacuum takes the form,
\begin{equation}
P_+=-\frac{3 b^2 M}{4 \pp  \left(b^2+r_+^2\right)^{5/2}}.
\end{equation}
In the 1LM the enthalpy and the internal energy equal
\begin{equation}
H=\frac{r_+^3}{2 \left(b^2+r_+^2\right)},\qquad
U=\frac{M r_+^3}{\left(b^2+r_+^2\right){}^{3/2}}=\frac{r_+}{2}.
\end{equation}

(iv) Model in Ref.~\cite{Balart:2014cga}.
This model is generated by a vacuum-like medium. The shape function takes the form,
\begin{equation}
f(r) = 1- \frac{2M}{r} \exp\left({-\frac{q^2}{2Mr}}\right),
\end{equation}
where $q$ stands for electric charge.
The Smarr formula is
\begin{equation}
M=\frac{A \kappa }{4 \pp }
+ \left(1-\me^{-\frac{q^2}{2 M r_+}}\right)M
+\frac{q^2 }{2 r_+}\me^{-\frac{q^2}{2 M r_+}},
\end{equation}
where $\epsilon$ and $P_+$ read
\begin{equation}
\epsilon=1-\me^{-\frac{q^2}{2 M r_+}},\qquad
P_+=-\frac{q^2}{8 \pp  r_+^4} \me^{-\frac{q^2}{2 M r_+}}.
\end{equation}
The enthalpy and internal energy are
\begin{equation}
H=\frac{r_+}{2}-\frac{ r_+}{6} \ln \left(\frac{2 M}{r_+}\right),\qquad
U=M \me^{-\frac{q^2}{2 M r_+}}=\frac{r_+}{2}.
\end{equation}

For the above four RBHs, we emphasize that
their internal energies calculated by Eq.\ \eqref{eq:internal-energy} equal $r_+/2$. 
This is not a coincidence.  
In fact, Eq.~\eqref{eq:internal-energy} gives exactly $r_+/2$ for all spherically symmetric RBHs with the shape function Eq.\ \eqref{eq:shape-function}.

An interesting property noted in Ref.~\cite{Azreg-Ainou:2014twa} is
that the $\Phi$-$Q$ term in Reissner-N\"ordstrom (RN) BHs
can be absorbed into $V$-$P_{+}$ term,
such that $\kappa \dif A/(8\pp)=\dif M - \Phi \dif Q$ becomes
Eq.~\eqref{eq:dymnikova-firstlaw2}.
This result implies that the Gliner vacuum provides a unified treatment for both RBHs
and SBHs in the establishment of 1LM.
In this way, the enthalpy and internal energy of RN BHs take the form,
\begin{equation}
H=\frac{r_+}{2}-\frac{Q^2}{6 r_+},\qquad
U=\frac{r_+}{2}.
\end{equation}
When $Q\to 0$, the enthalpy of RN BHs reduces to that of Schwarzschild BHs,  $H=r_+/2=M$,
which coincides with its internal energy.
For RN-AdS BHs, one has $H=U-Q^2/(6 r_+)- \Lambda  r_+^3/6$ with $U=r_+/2$.
It is worth emphasizing that the internal energies of SBHs, such as the Schwarzschild BHs, RN BHs, and  RN-AdS BHs we just mentioned, are calculated from the 1TL  but not from Eq.\ \eqref{eq:internal-energy}. The reason is that the energy density of SBHs is singular or vanishing at $r=0$, which leads to divergence of Eq.\ \eqref{eq:internal-energy} or makes the integration trivial.
In order to make Eq.\ \eqref{eq:internal-energy} suitable for SBHs, 
we introduce a cutoff of the radial coordinate for the SBHs whose energy densities are singular,
such that the integration is regularized.
Taking the RN BHs as an example, the regularized integration of internal energy is
\begin{equation}
U=\int^{r_{\rm +}}_{r_0} \int^{2\pp}_0\int^\pi_0\sqrt{-g} \rho\dif r \dif\phi\dif\theta,
\qquad
\rho=\frac{Q^2}{8 \pp  r^4},
\end{equation}
where $r_0$ is the cutoff. Since $U=r_+/2$, we can fix $r_0=Q^2/(2M)$.
Meanwhile, we can prove $r_0<r_-$, i.e.,
\begin{equation}
\label{eq:rn-critical}
Q^2/(2M)<M-\sqrt{M^2-Q^2},
\end{equation}
which implies that the cutoff is located inside the inner horizon.
As to the Schwarzschild BHs,
one can use the interior  metric of Schwarzschild BHs~\cite{Schwarzschild:1916ae} 
and suppose that the radius of the matter surface equals $2M$,
thus one can derive the energy density,
\begin{equation}
\rho(r)= \frac{3}{32 \pp  M^2} \left[\mathcal{H} \left(\frac{r}{2 M}\right)-\mathcal{H}\left(\frac{r}{2 M}-1\right)
\right],
\end{equation}
where $\mathcal{H} (\cdot)$ is the Heaviside function. The energy in the volume $V=4\pp r_+^2/3$ takes the value $U=r_+/2=M$.

Now we summarize the features of the new pressure we have proposed in Models I \& II and also applied to the other four models.
At first, the Gliner pressure is not universal, i.e., it has different values for different models, which is similar to the Hawking temperature that is horizon dependent. Next, there are no conceptional contradictions between the pressure and the vacuum or vacuum-like matter that generates RBHs, i.e.,
it is widely known that the RBHs are generated by the vacuum or vacuum-like matter with negative pressure rather than
the AdS constant with positive pressure. At last, we find the following relation,
\begin{equation}
|P_+|\propto R(0),
\end{equation}
where $R(0)$ can be explained as the average energy of a vacuum.
This gives rise to our conclusion that the new pressure we proposed, i.e., the radial pressure of the Gliner vacuum originates from the average energy of the Gliner vacuum, which is consistent with the case for SBHs where the pressure comes simply from the average energy of (AdS) vacuum.
In summary, by introducing the Gliner pressure as the pressure of RBHs we avoid such a contradiction that a RBH generated by the matter (vacuum) with negative pressure has positive pressure because this contradiction is physically unacceptable.

\section{Ruppeiner geometry of regular black holes}
\label{sec:ruppeiner}

Having the reconstructed 1LT, see Eq.~(\ref{lawthermo}), at hand,
we apply the Gibbs energy, $G=H-TS$, as the starting point
and calculate the line element by following Ref.~\cite{Wei:2019yvs},
we arrived at
\begin{equation}
\dif l^2 =\frac{1}{k_{\rm B} }
\left[
\frac{C_{P_{+}}}{T^2}\dif  T^2
+\frac{1}{T}
\left(\frac{\partial V}{\partial P_{+}}\right)_T
\dif P_{+}^2
\right],
\end{equation}
where $k_{\rm B}$ is Boltzmann constant and the capacity
at constant pressure is defined as usual,
\begin{equation}
C_{P_{+}}=T\left(\frac{\partial S}{\partial T}\right)_{P_{+}}.
\end{equation}
Considering the metric with the shape function, Eq.~\eqref{eq:shape-function},
we find
\begin{equation}
P_{+}= -\frac{M \sigma'(r_+)}{ 4\pp r_+^2},\qquad
T=\frac{1}{4\pp r_+}-\frac{M \sigma'(r_+)}{ 2\pp r_+},
\end{equation}
and then obtain the equation of state by combining the above two formulas,
\begin{equation}
P_{+}= \frac{T}{2 r_+}
-\frac{1}{8 \pp  r_+^2},
\end{equation}
whose dependence on parameters,
such as mass, charge, etc., is hidden in the outer horizon.
Based on this equation of state,  we derive the thermodynamic curvature for a general spherically symmetric RBH with Eq.~\eqref{eq:shape-function} as its shape function,
\begin{equation}
\label{eq:thermal-curvature}
\frac{\cal R}{2\pp T^2}=
y \left(\sqrt{1-y}-3\right)-4 \left(\sqrt{1-y}-1\right),
\end{equation}
where $y$ defined by $y\equiv 2P_{+}/(\pp T^2)$ is dimensionless.
Since $P_{+}< 0$, we deduce ${\cal R}/T^2< 0$. That is to say,
the interaction of all spherical symmetric RBHs with a single shape function is attractive from the microscopic perspective.
Nevertheless, it is known that RBHs are generated by matters which are of repulsive interaction around the center,
thus a natural problem is how a repulsive matter forms an attractive black hole.
Let us give a quantitative analysis in terms of the strong energy condition (SEC) of RBHs.

\section{Repulsive and attractive interactions inside and outside regular black holes}
\label{sec:strong-energy}

Instead of analyzing the energy-momentum tensor of matter $T^{\mu}_{\;\nu}$,
we concentrate on the diagonalized Einstein tensor,
\begin{equation}
G^{\mu}_{\;\nu}=\diag\left\{
-\frac{2 M \sigma '}{r^2},-\frac{2 M \sigma '}{r^2},-\frac{M \sigma ''}{r},-\frac{M \sigma ''}{r}
\right\},
\end{equation}
because $G^{\mu}_{\;\nu}$ and $T^{\mu}_{\;\nu}$ are equivalent due to the Einstein equation $G^{\mu}_{\;\nu}=8\pp T^{\mu}_{\;\nu}$.
The energy density $\rho$ and pressures $\rho_r$ and $\rho_\perp$ can then be obtained,
\begin{equation}
\label{eq:therm-variables}
\rho=\frac{ M \sigma '}{4\pp r^2},\qquad
p_r=-\frac{ M \sigma '}{4\pp r^2},\qquad
p_\perp=-\frac{M \sigma ''}{8\pp r}.
\end{equation}
Moreover, in order to discuss the SEC,  we introduce parameter  $\xi$,
\begin{equation}
\label{eq:beta}
\gamma \equiv \rho+p_r+2p_\perp=-\frac{M \sigma ''}{4 \pp  r},
\end{equation}
whose sign indicates the attractive ($\gamma>0$) or repulsive ($\gamma<0$) interaction.
This can be understood clearly from Raychaudhuri's equation~\cite{Carroll:2004st}.
Moreover, if the expansion, rotation, and shear
terms in Raychaudhuri's equation are neglected when compared with the variation of expansion,  one has
	\begin{equation}
		\frac{\dif \xi}{\dif \tau} =-4\pp \gamma,
	\end{equation}
where $\xi$ denotes the expansion of geodesics and $\tau$ affine parameter. Based on this equation, one can determine that the gravity is attractive ($\gamma>0$) or repulsive ($\gamma<0$).

Further, since $\sigma\sim r^n$ with $n\ge 3$ as $r\to 0$, see App.\ \ref{app:asym-sigma} for the details, we can deduce that $\sigma''>0$ and $\gamma<0$ around $r=0$, namely, the matters generating RBHs are of repulsive interaction, which is also known as the violation of SEC~\cite{Ansoldi:2008jw}.
Meanwhile, there is a special point arising from
$\sigma''(r_*)=0$, where $r_*$ can be regarded as the point of phase transition.
This point is special because the matters generating RBHs are of repulsive interaction in the range of $0<r<r_*$,
while they are of attractive interaction in the range of  $r>r_*$, i.e., $r_*$ separates the two phases of RBHs.

Now let us estimate the position of $r_*$.
At first, we note that Eq.\ \eqref{eq:beta} can be regarded as Newton's equation of a one-dimensional particle with mass $M$, i.e.,
\begin{equation}
-\Phi' = M \sigma'',\qquad
\Phi'\equiv 4\pp r \gamma,
\end{equation}
where $\Phi$ is ``potential'' and can be solved analytically,
\begin{equation}
\Phi=-M \sigma'+\Phi_0.
\end{equation}
Here $\Phi_0$ is an integration constant and $\Phi-\Phi_0<0$ in $r\in (0,\infty)$
because $\sigma$ is a monotone increasing function of $r$.
Then, considering the asymptotic behaviors of $\sigma$ at $r=0$ and $r\to\infty$,
we obtain
\begin{equation}
\lim_{r\to0}  \sigma'=0,\qquad
\lim_{r\to\infty}  \sigma'=0,
\end{equation}
which implies that
$\Phi$ is a potential well with one global minimum at $r=r_*$. The reason is that $\sigma$ is a sigmoid function and thus its first derivative is bell shaped. In other words, we have
$\sigma'(r_*)>\sigma'(r_{\rm ext})=1/(2M)$, where $\sigma'(r_{\rm ext})=1/(2M)$ comes from the combination of $T(r_{\rm ext})=0$ and $f(r_{\rm ext})=0$,
but we still cannot determine whether $r_*>r_{\rm ext}$ or $r_*<r_{\rm ext}$,
where $r_{\rm ext}$ denotes the  radius of extreme  RBHs.
At last, we know that the temperature of a RBH is nonnegative and vanishes at $r=r_{\rm ext}$,
from which we can deduce $T'(r_{\rm ext})>0$, i.e.,
\begin{equation}
\sigma ''\left(r_{\rm ext}\right)<\frac{2 \sigma '\left(r_{\rm ext}\right)}{r_{\rm ext}}-\frac{2 \sigma \left(r_{\rm ext}\right)}{r_{\rm ext}^2},
\end{equation}
then applying $\sigma '\left(r_{\rm ext}\right)=1/(2 M)$ and $\sigma \left(r_{\rm ext}\right)=r_{\rm ext}/(2 M)$ to replace $\sigma '\left(r_{\rm ext}\right)$ and $\sigma\left(r_{\rm ext}\right)$, we arrive at
\begin{equation}
\sigma ''\left(r_{\rm ext}\right)<0,
\end{equation}
which helps us rule out $r_*>r_{\rm ext}$.

In summary, we have $r_*<r_{\rm ext}$, i.e., the point of phase transition  should be located inside the extreme horizon,
which gives us an explanation of how a repulsive matter forms an attractive black hole.
The whole physical picture should be like this:
Along the radial coordinate,
the matter first shows the repulsive interaction around $r=0$; when $r$ passes through the phase transition point $r_*$,
the repulsive interaction becomes the attractive one.
Since thermodynamics describes a BH as a quantum system from the outside,
the interaction of RBHs should reflect the attractive nature. Here we have explained how a repulsive matter forms a black hole with attractive interaction.
In addition, we note from Eq.\ \eqref{eq:sigma-pp},
\begin{equation}
\gamma\propto-\left(R-\sqrt{2E}\right),
\end{equation}
which implies that the sign of $\gamma$ depends on the competition between two scalar curvatures $R$ and $E$ outside  RBHs. When $R<\sqrt{2E}$, the interaction is attractive, while $R>\sqrt{2E}$ means a repulsive interaction.
The balance $R=\sqrt{2E}$ corresponds to the phase transition point $r_*$.

\section{Comparison with Reissner-Nordstr\"om black hole}
\label{sec:RN-RBH}

To give a more intuitive illustration of the interaction structure of RBHs,  in this section we compare them with Reissner-Nordstr\"om black hole (RN BH) by considering the full version of the Raychaudhuri equation.
Because RN BH has also attractive interaction outside the horizon and repulsive one in the vicinity of $r=0$
\cite{Maluf:2014nsa,Ong:2020xwv}.

According to our formula Eq.\ \eqref{eq:thermal-curvature} for thermodynamic curvature of a general RBH with spherical
symmetry, RN BH has a negative thermodynamic curvature as well, i.e.\ attractive interaction outside the horizon, but its local structure of interaction in the vicinity of $r=0$ is different from the RBHs'.

To show the difference,
we start with the {\em ingoing} radial  geodesics on a general spherically symmetric metric with single shape function Eq.\ \eqref{eq:shape-function}, 
the tangent vector field of those geodesics is
\begin{equation}
    u_\alpha=\left(
    -1, u_r, 0,0 
    \right),
\end{equation}
where
\begin{equation}
    u_r=-f^{-1}\sqrt{2M\sigma(r)/r}.
\end{equation}
The main element in the Raychaudhuri equation is a B-tensor, which is defined as the divergence of the tangent vector, 
$B_{\alpha\beta}\coloneqq\nabla_{\beta} u_\alpha$, thus the expansion scalar $\xi$ 
can be expressed by the trace of $B_{\alpha\beta}$,
$\xi=\Tr B_{\alpha\beta}$. 
In our case, it reads
\begin{equation}
 \xi=-\sqrt{\frac{M }{2 r^3 \sigma }}
 \left(r \sigma '+3 \sigma \right).
\end{equation}
Thus the expansion scalar is negative definite if $\sigma$ is a monotonically non-decreasing function of $r$.
The general Raychaudhuri equation in our notation is
\begin{equation}
		\frac{\dif \xi}{\dif \tau}
		=\Theta,\quad
		\Theta\coloneqq
		-B_{\alpha\beta}B^{\alpha\beta}-4\pp \gamma,
\end{equation}
where
\begin{equation}
   B_{\alpha\beta}B^{\alpha\beta}= \frac{M }{2 r^3 \sigma }
   \left[r^2 (\sigma ')^2-2 r \sigma  \sigma '+9 \sigma^2\right].
\end{equation}

For RN BN, its $\sigma$ according to the Eq.\ \eqref{eq:shape-function} is
\begin{equation}
    \sigma_{\rm rn}=1-\frac{Q^2}{2 M r}.
\end{equation}
The $\Theta$ is then can be computed
\begin{equation}
   \Theta_{\rm rn}
   =-\frac{9 \Delta ^2+2 \Delta  Q^2+Q^4}{4 \Delta  r^4}\sim \frac{2 Q^2}{r^4}+O\left(\frac{1}{r^3}\right),
\end{equation}
where $\Delta =2 M r-Q^2$. 
It notes that this quantity is not positive definite and divergent as $r$ approaches to zero. 
Moreover, there exist two phases
\begin{subequations}
\begin{equation}
\label{eq:phase-1}
    \Theta_{\rm rn}<0,\quad 
    \text{when}\; r>Q^2/(2 M),
\end{equation}
\begin{equation}
\label{eq:phase-2}
    \Theta_{\rm rn}>0,\quad 
    \text{when}\; r<Q^2/(2 M).
\end{equation}
\end{subequations}
where the critical point $r_0=Q^2/(2M)$ is located inside the inner horizon because of Eq.\ \eqref{eq:rn-critical}.
In other words, the change of expansion scalar is negative in the first phase Eq.\ \eqref{eq:phase-1}, while becomes positive when the geodesics cross the critical point $r_0$ into the second phase Eq.\ \eqref{eq:phase-2}.
Finally, the SEC of RN BH holds in the whole domain
\begin{equation}
   \gamma_{\rm rn}=\frac{Q^2}{4 \pp r^4}>0.
\end{equation}
This aspect is consistent with the result 
obtained from the weak-field approximation \cite{Ong:2020xwv}, i.e., 
The RN BH reveals repulsive interaction in the vicinity of $r=0$, although the SEC holds everywhere.

The situation of RBHs is different. 
Let us take the Bardeen BH as an example.
Its $\Theta$ is strictly negative, i.e.,
it has only one phase compared with the RN BH.
\begin{equation}
    \Theta_{\rm b}
   =-\frac{3 M r^2 \left(10 b^2+3 r^2\right)}{2 \left(b^2+r^2\right)^{7/2}}<0,
\end{equation}  
and the change of expansion scalar vanishes at the center of BH because
\begin{equation}
    \Theta_{\rm b}\sim -\frac{15 M r^2}{b^5}+O\left(r^3\right),
\end{equation}
In fact, for all RBHs, if $\sigma$ has power expansion at $r=0$, Eq.\
\eqref{eq:series-sigma}
then round BH center one has
\begin{equation}
  \Theta=  4 a_{1} M r+\left(10 a_{2}-\frac{a_{1}^2}{2 a_{0}}\right) M r^2+O\left(r^3\right),
\end{equation}
where $a_i$ are abbreviate notation of the coefficients of series Eq.\
\eqref{eq:series-sigma}.
In other words, the expansion scalar stops changing at $r=0$. 
Oppositely, the change of expansion scalar for BH with a singularity at the center is always divergent.

The SEC of Bardeen BH does hold everywhere since we have
\begin{equation}
   \gamma_{\rm b}=\frac{3 M b^2  \left(3 r^2-2 b^2\right)}{4 \pp  \left(b^2+r^2\right)^{7/2}}.
\end{equation}
The SEC breaks when $r<b\sqrt{2/3}$, which is different from the situation of RN BH.

\section{Summary}
\label{sec:summary}

Starting from the idea of the Gliner vacuum,
we apply the approach given in Ref.~\cite{Azreg-Ainou:2014twa} to remove
the deformations in the 1LM for RBHs.
In addition, we provide a possible explanation for the deformation of the mass term.
The new 1LM redisplays the resemblance between RBHs and
traditional thermodynamic systems.
In other words, all the variables in the new 1LM have their thermodynamic counterparts,
in particular, the area law, $S\propto A$, is recovered.
Based on the reconstructed 1LM,
we give a self-consistent theory of Ruppeiner geometry,
and show that all RBHs with spherical symmetry and a single shape function
should have an attractive interaction in the range of $r\ge r_{\rm ext}>r_*$ from the microscopic perspective. This shows that our analyses of interactions inside and outside RBHs are consistent with the Ruppeiner thermodynamic geometry we established in Sec.\ \ref{sec:ruppeiner}.
Furthermore, the new 1LM offers a universal
treatment for both RBHs and SBHs.
However, the local properties of the interaction structures in the vicinity of $r=0$ between RBHs and SBHs are different.
Because of disappearing the center singularities, the expansion scalar of RBHs stops changing at $r=0$, 
while the change of expansion scalar for SBHs blows up. 
Finally, the explanation is given on how a repulsive matter forms a RBH with attractive interaction.
Our result may shed light on solving the related problems in superradiance and
area spectrum for RBHs.

\section*{Acknowledgments}

 C. Lan would like to acknowledge H. Geng for her useful discussions and constant support.
The authors are also grateful to H. Yang and X.-C. Cai for valuable remarks and suggestions.
The authors also would like to thank the anonymous referee for the helpful comments that improve
this work greatly.
This work was supported in part by the National Natural Science Foundation of China under Grants No.\ 11675081 and No.\ 12175108.

\appendix

\section{The representation of shape function via curvature invariants}
\label{app:shape-fun}

We start with the curvature invariants that are expressed \cite{Balart:2014cga} by the shape function, Eq.~(\ref{eq:shape-function}),
\begin{eqnarray}\label{eq:syst}
R &=&\frac{2M}{r^2}\left(2 \sigma'+r \sigma''\right),\\
R_2 & =&\frac{2M^2}{r^4}\left[4 (\sigma')^2+r^2 (\sigma'')^2\right],\\
K &=&\frac{4M^2}{r^6}\Big\{4\left[3\sigma^2-4r \sigma \sigma'+2r^2 (\sigma')^2\right]+4r^2\left(\sigma-r \sigma'\right) \sigma''+r^4 (\sigma'')^2\Big\}.
\end{eqnarray}
Although these equations are derived from a RBH,
they are valid for all spherically symmetric BHs with a single shape function.
We note that $R$, $R_2$, and $K$ contain $\sigma$ and its first and second derivatives with respect to $r$. Thus,
we can solve these three algebraic equations and express $\sigma$ and its derivatives in terms of the curvature invariants.
By ignoring redundant roots,\footnote{There are four roots originally, two of them are removed by the weak or null energy condition, $r \sigma ''\leq 2 \sigma '$.} we obtain
\begin{eqnarray}
\label{eq:fir-gr}
\sigma&=& \frac{r^3 }{24 M}\left(R\pm 2 \sqrt{3 K+R^2-6 R_2}+3 \sqrt{4 R_2-R^2}\right),\nonumber \\
\sigma'&=&\frac{r^2 }{8 M} \left(R + \sqrt{4 R_2-R^2}\right),\nonumber \\
\sigma''&=& \frac{r }{4 M}\left(R-\sqrt{4 R_2-R^2}\right),
\end{eqnarray}
where the different signs in $\sigma$ correspond to two regions separated by the line $r^2 \sigma''+6 \sigma =4 r \sigma '$ in the parameter space. The plus sign depicts the region $r^2 \sigma ''+6 \sigma > 4 r \sigma '$, while the minus one  $r^2 \sigma ''+6 \sigma < 4 r \sigma '$ is more closer to the center.
On the other side, the Riemann tensor breaks down into three parts in terms of the Ricci decomposition \cite{Weinberg:1972kfs,Wald:1984rg},
$R_{\mu\nu\rho\sigma}=W_{\mu\nu\rho\sigma}+S_{\mu\nu\rho\sigma}+E_{\mu\nu\rho\sigma}$,
where
$W_{\mu\nu\rho\sigma}$ is traceless part called the Weyl tensor, $S_{\mu\nu\rho\sigma}$ is scalar part,
and $E_{\mu\nu\rho\sigma}$ is semi-traceless part.
Moreover, we have the following relationships,
\begin{equation}
E\equiv E_{\mu\nu\rho\sigma}E^{\mu\nu\rho\sigma}=2 R_2-\frac{R^2}{2},\qquad
W\equiv W_{\mu\nu\rho\sigma}W^{\mu\nu\rho\sigma}=K - 2 R_2+\frac{R^2}{3}.
\end{equation}
Then applying these relations to replace $R_2$ and $K$ in Eq.~\eqref{eq:fir-gr},
we obtain
\begin{eqnarray}
\sigma&=&\frac{r^3}{24M} \left(R+3\sqrt{2E}\pm2\sqrt{3W}\right),\\
\sigma'&=&\frac{r^2}{8M} \left(R+\sqrt{2E}\right),\\
\sigma''&=&\frac{r}{4M} \left(R-\sqrt{2E}\right)\label{eq:sigma-pp}.
\end{eqnarray}

We can also express the Hawking temperature and Gliner pressure in terms of the  curvature invariants,
\begin{equation}
T=\frac{r }{24 \pp }\left(\sqrt{3W}-R\right)\Big|_{r=r_+},\qquad
P_{+}=-\frac{1}{32\pp}\left(\sqrt{2E}+R\right)\Big|_{r=r_+},
\end{equation}
where the case of the minus sign before $\sqrt{3W}$ in $T$ has been ruled out due to the positivity of temperature.
The zero point of $T$ as a function of $r_+$ corresponds to the solution of the algebraic equation $\sqrt{3W}=R$.
In other words, $\sqrt{3W}=R$ signifies the ground state of BH configurations.

\section{The asymptotic behavior of $\sigma$ around $r=0$}
\label{app:asym-sigma}

Using Eq.~\eqref{eq:shape-function}, we compute the scalar curvature,
\begin{equation}
\label{eq:ricci-diff}
R(r)=\frac{2 M \left(2 \sigma '+r \sigma ''\right)}{r^2}.
\end{equation}
If $R(r)$ is finite around $r=0$, its Taylor expansion has the form,
\begin{equation}
R(r)=\sum _{n=0}^{\infty } \frac{r^n R^{(n)}(0)}{n!}.
\end{equation}
By solving the ordinary differential equation, Eq.~\eqref{eq:ricci-diff}, and using the above Taylor expansion, we obtain a general solution for $\sigma$,
\begin{equation}
\tilde \sigma = c_1 + \frac{c_2}{r}
+r^3 \sum _{n=0}^{\infty } \frac{r^n R^{(n)}(0)}{2 M (n+3) (n+4) n!},\label{odesig}
\end{equation}
where $c_1$ and $c_2$ are two integration constants.
Substituting Eq.~(\ref{odesig}) into the Kretschmann scalar,
we can separate the scalar into two parts, where one is finite and the other divergent at $r=0$,
\begin{equation}
K=K^{\rm fin} + K^{\rm div},
\end{equation}
with
\begin{eqnarray}
K^{\rm fin} &=&
4 \sum _{n=0}^{\infty } \frac{r^n R^{(n)}(0)}{(n+3) (n+4) n!}
\sum _{m=0}^{\infty } \frac{\left(m^2+7 m+6\right) r^m R^{(m)}(0)}{(m+3) (m+4) m!}\nonumber \\
& &+4 \sum _{m=0}^{\infty } \frac{m (2 m+3) r^m R^{(m)}(0)}{(m+3) (m+4) m!}
\sum _{n=0}^{\infty } \frac{n r^n R^{(n)}(0)}{(n+3) (n+4) n!}
+r^2 \left(\sum _{n=0}^{\infty } \frac{(n-1) n r^n R^{(n)}(0)}{(n+3) (n+4) n!}\right)^2,
\end{eqnarray}
and
\begin{eqnarray}
 K^{\rm div} &=&
 \frac{8 c_2 M }{r^3}
 \sum _{n=0}^{\infty } \frac{n (n+1) r^n R^{(n)}(0)}{(n+3) (n+4) n!}+\frac{8 c_1 M }{r^4}
\sum _{n=0}^{\infty } \frac{n (3 n+5) r^n R^{(n)}(0)}{(n+3) (n+4) n!}\nonumber \\
 & &+\frac{192 c_1 c_2 M^2}{r^7}+\frac{48 c_2^2 M^2}{r^6}+\frac{224 c_1^2 M^2}{r^8}.
\end{eqnarray}
A regular black hole implies $K^{\rm div}=0$, i.e.,  $c_1=0=c_2$.
As a result, we derive the asymptotic behavior of the Kretschmann scalar,
\begin{equation}
K=K^{\rm fin}\sim
\frac{\left[R(0)\right]^2}{6}+
\frac{R(0) R'(0)}{3} r +O(r^2),
\end{equation}
and the asymptotic behavior of $\sigma$ around $r=0$,
\begin{equation}
\label{eq:series-sigma}
\sigma = r^3 \sum _{n=0}^{\infty } \frac{r^n R^{(n)}(0)}{2 M (n+3) (n+4) n!}
\sim \frac{r^3 R(0)}{24 M}+\frac{r^4 R'(0)}{40 M}+O(r^5).
\end{equation}


\bibliographystyle{utphys}

\bibliography{references}

\providecommand{\href}[2]{#2}\begingroup\raggedright\begin{thebibliography}{10}

\bibitem{Ruppeiner:1995zz}
G.~Ruppeiner, ``{Riemannian geometry in thermodynamic fluctuation theory},''
  \href{http://dx.doi.org/10.1103/RevModPhys.67.605}{{\em Rev. Mod. Phys.}
  {\bfseries 67} (1995) 605--659}. [Erratum: Rev.Mod.Phys. 68, 313--313
  (1996)].

\bibitem{Wei:2019uqg}
S.-W. Wei, Y.-X. Liu, and R.~B. Mann, ``{Repulsive interactions and universal
  properties of charged Anti\textendash{}de Sitter black hole
  microstructures},''
  \href{http://dx.doi.org/10.1103/PhysRevLett.123.071103}{{\em Phys. Rev.
  Lett.} {\bfseries 123} no.~7, (2019) 071103},
  \href{http://arxiv.org/abs/1906.10840}{{\ttfamily arXiv:1906.10840 [gr-qc]}}.

\bibitem{Hawking:1974sw}
S.~W. Hawking, ``Particle creation by black holes,''
  \href{http://dx.doi.org/10.1007/BF02345020}{{\em Commun. Math. Phys.}
  {\bfseries 43} no.~3, (Aug., 1975) 199--220}. [Erratum: Commun.Math.Phys. 46,
  206 (1976)].

\bibitem{Bekenstein:1973ur}
J.~D. Bekenstein, ``{Black holes and entropy},''
  \href{http://dx.doi.org/10.1103/PhysRevD.7.2333}{{\em Phys. Rev. D}
  {\bfseries 7} (1973) 2333--2346}.

\bibitem{Hawking:1971tu}
S.~W. Hawking, ``{Gravitational radiation from colliding black holes},''
  \href{http://dx.doi.org/10.1103/PhysRevLett.26.1344}{{\em Phys. Rev. Lett.}
  {\bfseries 26} (1971) 1344--1346}.

\bibitem{Gibbons:1976ue}
G.~W. Gibbons and S.~W. Hawking, ``Action integrals and partition functions in
  quantum gravity,'' \href{http://dx.doi.org/10.1103/PhysRevD.15.2752}{{\em
  Phys. Rev. D} {\bfseries 15} (May, 1977) 2752--2756}.

\bibitem{Wald:1993nt}
R.~M. Wald, ``Black hole entropy is the noether charge,''
  \href{http://dx.doi.org/10.1103/PhysRevD.48.R3427}{{\em Phys. Rev. D}
  {\bfseries 48} (Oct., 1993) R3427--R3431},
  \href{http://arxiv.org/abs/gr-qc/9307038}{{\ttfamily arXiv:gr-qc/9307038
  [gr-qc]}}.

\bibitem{Brustein:2007jj}
R.~Brustein, D.~Gorbonos, and M.~Hadad, ``Wald's entropy is equal to a quarter
  of the horizon area in units of the effective gravitational coupling,''
  \href{http://dx.doi.org/10.1103/PhysRevD.79.044025}{{\em Phys. Rev. D}
  {\bfseries 79} (Feb., 2009) 044025},
  \href{http://arxiv.org/abs/0712.3206}{{\ttfamily arXiv:0712.3206 [hep-th]}}.

\bibitem{Zhang:2016ilt}
Y.~Zhang and S.~Gao, ``First law and {Smarr} formula of black hole mechanics in
  nonlinear gauge theories,''
  \href{http://dx.doi.org/10.1088/1361-6382/aac9d4}{{\em Class. Quant. Grav.}
  {\bfseries 35} no.~14, (Oct., 2016) 145007},
  \href{http://arxiv.org/abs/1610.01237}{{\ttfamily arXiv:1610.01237 [gr-qc]}}.

\bibitem{Nicolini:2005vd}
P.~Nicolini, A.~Smailagic, and E.~Spallucci, ``Noncommutative geometry inspired
  {Schwarzschild} black hole,''
  \href{http://dx.doi.org/10.1016/j.physletb.2005.11.004}{{\em Phys. Lett. B}
  {\bfseries 632} no.~4, (Jan., 2006) 547--551},
  \href{http://arxiv.org/abs/gr-qc/0510112}{{\ttfamily arXiv:gr-qc/0510112
  [gr-qc]}}.

\bibitem{Dabholkar:2012zz}
A.~Dabholkar and S.~Nampuri, ``{Quantum black holes},''
  \href{http://dx.doi.org/10.1007/978-3-642-25947-0_5}{{\em Lect. Notes Phys.}
  {\bfseries 851} (2012) 165--232},
  \href{http://arxiv.org/abs/1208.4814}{{\ttfamily arXiv:1208.4814 [hep-th]}}.

\bibitem{Banerjee:2008gc}
R.~Banerjee, B.~R. Majhi, and S.~Samanta, ``{Noncommutative Black Hole
  Thermodynamics},'' \href{http://dx.doi.org/10.1103/PhysRevD.77.124035}{{\em
  Phys. Rev. D} {\bfseries 77} (2008) 124035},
  \href{http://arxiv.org/abs/0801.3583}{{\ttfamily arXiv:0801.3583 [hep-th]}}.

\bibitem{Gliner:1966}
E.~B. Gliner, ``{Algebraic Properties of the Energy-momentum Tensor and
  Vacuum-like States of Matter},'' {\em Sov. Phys. JETP} {\bfseries 22} (1966)
  378.

\bibitem{Ansoldi:2008jw}
S.~Ansoldi, ``{Spherical black holes with regular center: A Review of existing
  models including a recent realization with Gaussian sources},'' in {\em
  {Conference on Black Holes and Naked Singularities}}.
\newblock 2, 2008.
\newblock \href{http://arxiv.org/abs/0802.0330}{{\ttfamily arXiv:0802.0330
  [gr-qc]}}.

\bibitem{Silbergleit2017}
A.~S. Silbergleit and A.~D. Chernin, {\em Why Does the Universe Expand? (A
  Tribute to E.B. Gliner)},
  \href{http://dx.doi.org/10.1007/978-3-319-57538-4_6}{pp.~59--70}.
\newblock Springer International Publishing, Cham, 2017.
\newblock \url{https://doi.org/10.1007/978-3-319-57538-4_6}.

\bibitem{Dymnikova:1992ux}
I.~Dymnikova, ``{Vacuum nonsingular black hole},''
  \href{http://dx.doi.org/10.1007/BF00760226}{{\em Gen. Rel. Grav.} {\bfseries
  24} (1992) 235--242}.

\bibitem{Elizalde:2002yz}
E.~Elizalde and S.~R. Hildebrandt, ``{The Family of regular interiors for
  nonrotating black holes with $T^0_0=T^1_1$},''
  \href{http://dx.doi.org/10.1103/PhysRevD.65.124024}{{\em Phys. Rev. D}
  {\bfseries 65} (2002) 124024},
  \href{http://arxiv.org/abs/gr-qc/0202102}{{\ttfamily arXiv:gr-qc/0202102}}.

\bibitem{Petrov:2019}
A.~Z. Petrov, {\em {New methods in general theory of relativity (in Russian)}}.
\newblock URSS, Moscow, 2019.

\bibitem{Ma:2014qma}
M.-S. Ma and R.~Zhao, ``{Corrected form of the first law of thermodynamics for
  regular black holes},''
  \href{http://dx.doi.org/10.1088/0264-9381/31/24/245014}{{\em Class. Quant.
  Grav.} {\bfseries 31} (2014) 245014},
  \href{http://arxiv.org/abs/1411.0833}{{\ttfamily arXiv:1411.0833 [gr-qc]}}.

\bibitem{Azreg-Ainou:2014twa}
M.~Azreg-A\"\i{}nou, ``{Black hole thermodynamics: No inconsistency via the
  inclusion of the missing $P-V$ terms},''
  \href{http://dx.doi.org/10.1103/PhysRevD.91.064049}{{\em Phys. Rev. D}
  {\bfseries 91} (2015) 064049},
  \href{http://arxiv.org/abs/1411.2386}{{\ttfamily arXiv:1411.2386 [gr-qc]}}.

\bibitem{Weinberg:1972kfs}
S.~Weinberg, {\em {Gravitation and Cosmology}: {Principles and Applications of
  the General Theory of Relativity}}.
\newblock John Wiley and Sons, New York, 1972.

\bibitem{Melgarejo:2020mso}
G.~Melgarejo, E.~Contreras, and P.~Bargue\~no, ``{Regular black holes with
  exotic topologies},''
  \href{http://dx.doi.org/10.1016/j.dark.2020.100709}{{\em Phys. Dark Univ.}
  {\bfseries 30} (2020) 100709}.

\bibitem{Bargueno:2020ais}
P.~Bargue\~no, ``{Some global, analytical and topological properties of regular
  black holes},'' \href{http://dx.doi.org/10.1103/PhysRevD.102.104028}{{\em
  Phys. Rev. D} {\bfseries 102} no.~10, (2020) 104028},
  \href{http://arxiv.org/abs/2008.02680}{{\ttfamily arXiv:2008.02680 [gr-qc]}}.

\bibitem{Lan:2021klp}
C.~Lan and Y.-G. Miao, ``Entropy and topology of regular black holes,''
  \href{http://arxiv.org/abs/2105.00218}{{\ttfamily arXiv:2105.00218 [gr-qc]}}.

\bibitem{Landau:vol8}
L.~D. Landau and E.~M. Lifshits, {\em {Electrodynamics of continuous media}},
  vol.~8 of {\em Course of Theoretical Physics}.
\newblock Fizmatlit, Moscow, 2005.

\bibitem{Kastor:2009wy}
D.~Kastor, S.~Ray, and J.~Traschen, ``Enthalpy and the mechanics of {AdS} black
  holes,'' \href{http://dx.doi.org/10.1088/0264-9381/26/19/195011}{{\em Class.
  Quant. Grav.} {\bfseries 26} no.~19, (Apr., 2009) 195011},
  \href{http://arxiv.org/abs/0904.2765}{{\ttfamily arXiv:0904.2765 [hep-th]}}.

\bibitem{Dolan:2010ha}
B.~P. Dolan, ``The cosmological constant and black-hole thermodynamic
  potentials,'' \href{http://dx.doi.org/10.1088/0264-9381/28/12/125020}{{\em
  Class. Quant. Grav.} {\bfseries 28} no.~12, (May, 2011) 125020},
  \href{http://arxiv.org/abs/1008.5023}{{\ttfamily arXiv:1008.5023 [gr-qc]}}.

\bibitem{Sakharov:1966aja}
A.~D. Sakharov, ``{The initial stage of an expanding Universe and the
  appearance of a nonuniform distribution of matter},'' {\em Sov. Phys. JETP}
  {\bfseries 22} (1966) 241.

\bibitem{Bardeen:1973gs}
J.~M. Bardeen, B.~Carter, and S.~W. Hawking, ``The four laws of black hole
  mechanics,'' \href{http://dx.doi.org/10.1007/BF01645742}{{\em Commun. Math.
  Phys.} {\bfseries 31} no.~2, (June, 1973) 161--170}.

\bibitem{Gulin:2017ycu}
L.~Gulin and I.~Smoli\'c, ``Generalizations of the {Smarr} formula for black
  holes with nonlinear electromagnetic fields,''
  \href{http://dx.doi.org/10.1088/1361-6382/aa9dfd}{{\em Class. Quant. Grav.}
  {\bfseries 35} no.~2, (Oct., 2017) 025015},
  \href{http://arxiv.org/abs/1710.04660}{{\ttfamily arXiv:1710.04660 [gr-qc]}}.

\bibitem{Dymnikova:1998pm}
I.~Dymnikova and B.~Soltysek, ``{Spherically symmetric space-time with two
  cosmological constants},''
  \href{http://dx.doi.org/10.1023/A:1026619228583}{{\em Gen. Rel. Grav.}
  {\bfseries 30} (1998) 1775--1793}.

\bibitem{Hayward:2005gi}
S.~A. Hayward, ``Formation and evaporation of nonsingular black holes,''
  \href{http://dx.doi.org/10.1103/PhysRevLett.96.031103}{{\em Phys. Rev. Lett.}
  {\bfseries 96} (Jan., 2006) 031103},
  \href{http://arxiv.org/abs/gr-qc/0506126}{{\ttfamily arXiv:gr-qc/0506126
  [gr-qc]}}.

\bibitem{Bardeen:1968non}
J.~M. Bardeen, ``Non-singular general-relativistic gravitational collapse,'' in
  {\em Conference Proceedings of GR5, Tbilisi, USSR}, vol.~174.
\newblock 1968.

\bibitem{Balart:2014cga}
L.~Balart and E.~C. Vagenas, ``Regular black holes with a nonlinear
  electrodynamics source,''
  \href{http://dx.doi.org/10.1103/PhysRevD.90.124045}{{\em Phys. Rev. D}
  {\bfseries 90} (Dec., 2014) 124045},
  \href{http://arxiv.org/abs/1408.0306}{{\ttfamily arXiv:1408.0306 [gr-qc]}}.

\bibitem{Schwarzschild:1916ae}
K.~Schwarzschild, ``{On the gravitational field of a sphere of incompressible
  fluid according to Einstein's theory},'' {\em Sitzungsber. Preuss. Akad.
  Wiss. Berlin (Math. Phys. )} {\bfseries 1916} (1916) 424--434,
  \href{http://arxiv.org/abs/physics/9912033}{{\ttfamily
  arXiv:physics/9912033}}.

\bibitem{Wei:2019yvs}
S.-W. Wei, Y.-X. Liu, and R.~B. Mann, ``{Ruppeiner geometry, phase transitions,
  and the microstructure of charged AdS black holes},''
  \href{http://dx.doi.org/10.1103/PhysRevD.100.124033}{{\em Phys. Rev. D}
  {\bfseries 100} no.~12, (2019) 124033},
  \href{http://arxiv.org/abs/1909.03887}{{\ttfamily arXiv:1909.03887 [gr-qc]}}.

\bibitem{Carroll:2004st}
S.~M. Carroll, {\em {Spacetime and Geometry}}.
\newblock Cambridge University Press, 7, 2019.

\bibitem{Maluf:2014nsa}
J.~W. Maluf, ``{Repulsive gravity near naked singularities and point massive
  particles},'' \href{http://dx.doi.org/10.1007/s10714-014-1734-y}{{\em Gen.
  Rel. Grav.} {\bfseries 46} (2014) 1734},
  \href{http://arxiv.org/abs/1401.0741}{{\ttfamily arXiv:1401.0741 [gr-qc]}}.

\bibitem{Ong:2020xwv}
Y.~C. Ong, ``{Space\textendash{}time singularities and cosmic censorship
  conjecture: A Review with some thoughts},''
  \href{http://dx.doi.org/10.1142/S0217751X20300070}{{\em Int. J. Mod. Phys. A}
  {\bfseries 35} no.~14, (2020) 14},
  \href{http://arxiv.org/abs/2005.07032}{{\ttfamily arXiv:2005.07032 [gr-qc]}}.

\bibitem{Wald:1984rg}
R.~M. Wald,
  \href{http://dx.doi.org/10.7208/chicago/9780226870373.001.0001}{{\em General
  Relativity}}.
\newblock University of Chicago Press, Chicago, USA, 1984.

\end{thebibliography}\endgroup


\end{document}